\documentclass[twocolumn,prb,showpacs,preprintnumbers,amsmath,amssymb]{revtex4}
\usepackage{graphicx}
\usepackage{dcolumn}
\usepackage{bm}

\begin{document}


\title{Single spin optical \textit{read-out} in CdTe/ZnTe quantum dot studied by photon correlation spectroscopy}

\author{J. Suf\mbox{}fczy\'nski}
\email{Jan.Suffczynski@fuw.edu.pl}
\author{K. Kowalik}
\author{T. Kazimierczuk}
\author{A. Trajnerowicz}
\author{M. Goryca}
\author{P. Kossacki}
\author{A. Golnik}
\author{M. Nawrocki}
\author{J. A. Gaj}
\affiliation{Institute of Experimental Physics, University of Warsaw, Ho\.za 69, 00-681 Warsaw, Poland}%

\author{G. Karczewski}
\affiliation{Institute of Physics, Polish Academy of Sciences, Al. Lotnik\'ow 32/64, 02-668 Warsaw, Poland}%


\begin{abstract}
Spin dynamics of a single electron and an exciton confined in CdTe/ZnTe quantum dot
is investigated by polarization-resolved correlation spectroscopy. Spin memory effects extending over at least a few tens of nanoseconds have been directly observed in magnetic field and described
quantitatively in terms of a simple rate equation model. We demonstrate an
effective (68\%) all-optical \textit{read-out} of the single carrier spin state through probing
the degree of circular polarization of exciton emission after capture of an
oppositely charged carrier. The perturbation introduced by the pulsed optical excitation serving to study the spin dynamics has been found to be the main source of the polarization loss in the read-out process. In the limit of low laser power the \textit{read-out} efficiency extrapolates to a value close to 100\%. The measurements allowed us as well to determine neutral exciton spin relaxation time ranging from $3.4 \pm 0.1$~ns at $B$~=~0~T to $16 \pm 3$ ns at $B$~=~5~T.
\end{abstract}


\maketitle

\section{\label{sec:Intro}Introduction}
One of the consequences of energy quantization in semiconductor quantum
dots (QDs) is the suppression of spin relaxation of confined carriers and
excitons.~\cite{Paillard, Khaetskii} Recent experiments conducted on
ensemble of III-V QDs have demonstrated that electrons confined in the QDs
preserve their spin polarization over microsecond~\cite{Ikezawa, Akimov} or
even millisecond timescales.~\cite{Kroutvar} It has been also shown, that spin of
the electron confined in the QD can be effectively optically \textit{read}
and \textit{written}.~\cite{Kroutvar, Cortez, Ebbens, Young07} These features,
complementing the fact that individual QDs can be used as
non-classical light sources~\cite{MichlerScience, Moreau} make QDs very
attractive for implementation in the developing field of quantum
information,~\cite{BB84, Ekert} where polarization-encoded single photons
would be utilized. However, several difficulties need to be overcome in
order to achieve effective operation of quantum qubits based on single
QDs. One of them is native QD anisotropy, which does not influence the
spin state of a single electron, but determines the eigenfunctions of exciton
and induces linear polarization of its emission. Thus, even if the spin of
the carrier is effectively stored, it can not be effectively \textit{read}.
The polarization of exciton which is formed in the optical
\textit{read-out} of the carrier spin state~\cite{Kroutvar} will be
determined by the anisotropy. A lot of effort has been devoted to
development of fabrication technique enabling creation of QDs possessing no anisotropy,~\cite{Ellis} however no straightforward method has been established so far.

In this work single carrier spin memory effects are studied by correlation spectroscopy technique. Description
of the experimental results with a simple rate equation model allowed us
to determine the degree of the carrier spin polarization conserved in the
process of optical \textit{read-out}. We quantify the impact of biexciton
formation on the loss of the carrier polarization memory. We determine
also neutral exciton ({\it X}) spin relaxation time - serving as a one of the model
parameters.

The paper is organized as follows. Section~\ref{sec:Sample} provides
information on the sample studied and the experimental setup. Experimental
results are collected in Sec.~\ref{sec:Results}, which is divided in three
parts. A summary of standard cw microphotoluminescence ($\mu$-PL)
characterization of the sample (Sec.~\ref{sec:Results.1}) is followed by
results of polarized biexciton-exciton ({\it XX}-{\it X}) crosscorrelation measurements (Sec.~\ref{sec:Results.2}), supplying information on {\it X} spin relaxation time. Charged exciton-neutral exciton ({\it CX}-{\it X}) polarized
crosscorrelation experiments reveal single carrier spin memory through effective optical \textit{read-out} (Sec.~\ref{sec:Results.3}). Section~\ref{sec:Model} contains the
rate equation model description of the experimental data.

\section{\label{sec:Sample}Sample and Experimental Setup}
Detailed macro- and micro-PL characterization of the sample used in this
work has been presented in Refs.~\onlinecite{Kudelski, KowalikAPPol,
SuffPRB}. The sample contains a single layer of QDs, self assembled out of
two monolayers of CdTe, embedded between ZnTe barriers. Typical density of
the QDs is 10$^{12}$~cm$^{-2}$.~\cite{Karczewski}

The sample was mounted directly on the front surface of a mirror type
microscope objective~\cite{Jasny} (numerical aperture = 0.7, spatial
resolution $\sim$0.5~$\mu$m) and cooled down to T~=~1.7~K in a pumped helium
cryostat with a superconducting coil. Microphotoluminescence was excited non-resonantly (above the
barrier band gap) with short ($<$1~ps) Ti$^{3+}$:Al$_{2}$O$_{3}$ laser pulses, delivered every 6.6~ns
at wavelength of 400~nm (after frequency doubling). The excitation beam was linearly polarized.

Photon correlations were measured in a Hanbury-Brown and
Twiss~\cite{HBTwiss} (HBT) type setup with spectral filtering. PL signal arising from the sample was divided in two
beams on a polarizing beamsplitter (BS) and directed to the entrances of two
grating monochromators (spectral resolution 200~$\mu$eV). The
monochromators were tuned to pass photons from a single excitonic
transition, chosen independently on each spectrometer. The signals were
then detected by two avalanche photodiodes. The diodes were connected to \textit{start} and \textit{stop}
inputs of coincidence counting electronics producing histogram (4096
time bins of 146~ps each) of correlated counts versus time interval
separating photon detection on the first and on the second diode. Total temporal resolution of the setup is estimated at 1.1 ns.

Polarization optics (combinations of halfwave and quarterwave retarders with a
linear polarizer) implemented in the HBT setup enabled detection of the
second-order correlation function for four linear or four circular
polarization combinations.

\section{\label{sec:Results}Experimental results}
\subsection{\label{sec:Results.1}Sample characterization}
\begin{figure}
\includegraphics[width=77mm]{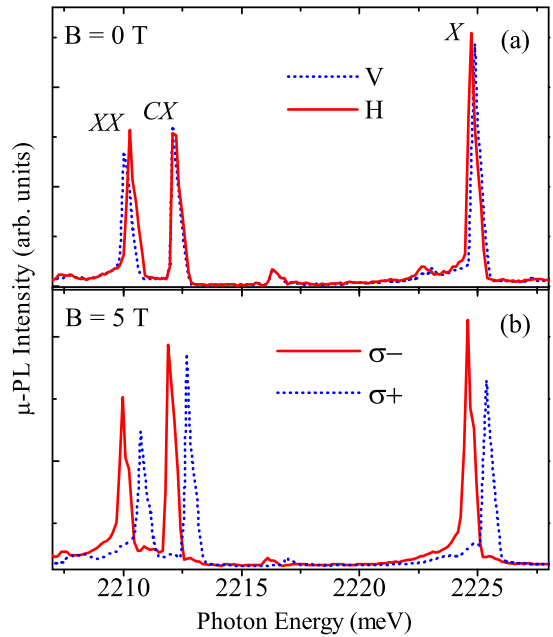}
\caption{\label{Fig:PLspectrum}(Color online) Polarization resolved
emission spectra of the selected quantum dot at $B$~=~0~T (a) and
$B$~=~5~T (b) detected in linear H-V and circular
$\sigma+$/$\sigma-$ polarization bases respectively. Excitation at
the energy of 3.1~eV with the average power $\sim$0.8~P$_{SAT}$ (the
saturation power of the X emission P$_{SAT}$~=1.2~$\mu$W at
$B$~=~0~T).}
\end{figure}

The $\mu$-PL from the QD layer covers the energy range between
2.20~eV and 2.32~eV. Excitonic transitions studied in this work were
selected from the low energy tail (2.20~eV -- 2.24~eV) of the
$\mu$-PL spectrum, since in this region lines of individual QDs are
well resolved and background counts are negligible.
Polarization resolved $\mu$-PL spectrum of the QD selected for this
study taken at $B$~=~0~T is presented in Fig.~\ref{Fig:PLspectrum}(a).
Dependence of integrated line intensities on excitation power
combined with auto- and crosscorrelation data allowed us to identify the
observed transitions as neutral exciton ({\it X}), charged exciton
({\it CX}) and biexciton ({\it XX}) recombination.~\cite{SuffPRB} As visible in
Fig.~\ref{Fig:PLspectrum}(a), {\it X} and {\it XX} lines exhibit anisotropic exchange
splitting (AES) in two linearly polarized components, resulting from
electron-hole exchange interaction in an anisotropic
QD.~\cite{BesombesPRL}

\begin{figure}
\includegraphics[width=79mm]{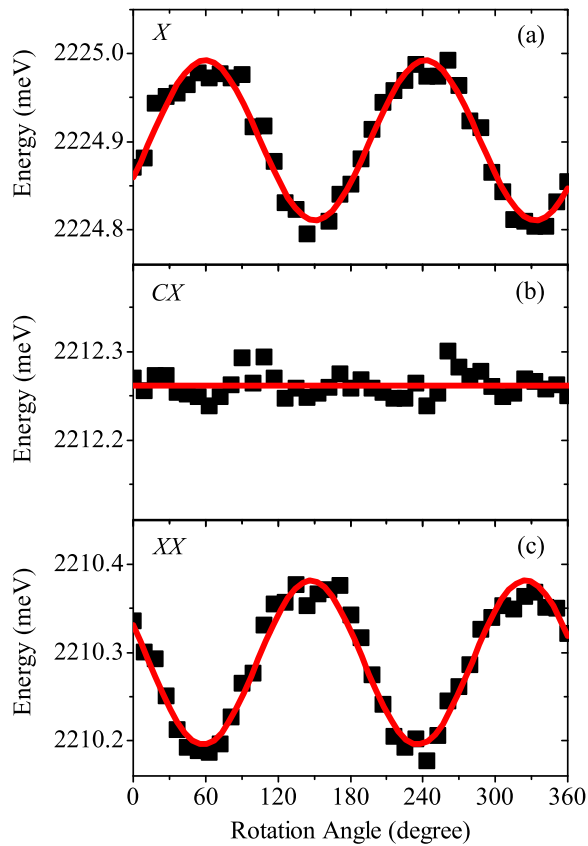}  %
\caption{\label{Fig:anis}(Color online) Emission energies of (a)
{\it X}, (b) {\it CX} and (c) {\it XX} versus angle of detected
linear polarization (points). Solid line represents a sinusoidal fit
(a) and (c), or a linear fit (b).}
\end{figure}

In order to determine the value of the AES and directions of linear
polarizations of the QD emission, energy positions of {\it X}, {\it CX} and {\it XX} as a function of detection polarization angle were measured (Fig.~\ref{Fig:anis}). No energy variation occurs in the case of {\it CX}, in agreement with the expectation (two identical carriers of the trion are in a singlet state). In the case of {\it X} and {\it XX}, oscillations of transition energy are observed with opposite phase and the same amplitude
for both lines. For the QD discussed below AES was determined to be $182 \pm 6$~$\mu$eV.

\begin{figure}
\includegraphics[width=75mm]{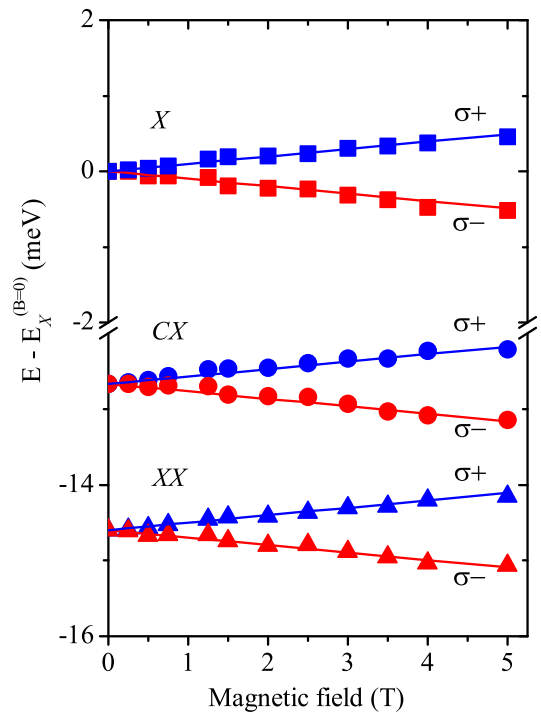}
\caption{\label{Fig:Zeeman}(Color online) Energy of {\it X}, {\it
CX} and {\it XX} emission relative to zero-field {\it X} energy,
measured in both circular polarizations as a function of magnetic
field in Faraday configuration. Solid lines represent the
calculation of the Zeeman splitting with Land\'e factor g~=~$-3.4$
the same for each line.}
\end{figure}

A common H-V basis of linear polarizations corresponding
to QD symmetry axes was determined as rotated 58$^\circ$~$\pm$~ 3$^\circ$
from laboratory axes. The QD symmetry axes do not correspond to the main
crystallographic axes of the sample, in agreement with previous anisotropy
measurements revealing random anisotropy orientation of CdTe/ZnTe
QDs.~\cite{KowalikAPPol}
Determination of the excitonic effective Land\'e factor based on Zeeman
splitting measurements (Fig.~\ref{Fig:Zeeman}) performed in magnetic field
ranging up to 5~T gave approximately the same value g~=~$-3.4~\pm~0.1$ for all
the three lines. This is expected since the same hole and electron g-factors contribute to the effective g-factor common for all the three excitonic complexes. The determined value is typical for QDs in the
investigated sample.~\cite{KowalikAPPol}
\subsection{\label{sec:Results.2}Polarized {\it XX}-{\it X} crosscorrelation measurements}
In this section, we present time and polarization resolved photon correlations
involving biexciton - exciton cascade. The measurements were performed in
magnetic field ranging up to 5~T and provided an estimate of the {\it X} spin
relaxation time. The spin relaxation time was found to increase with magnetic
field.

\begin{figure}
\includegraphics[width=78mm]{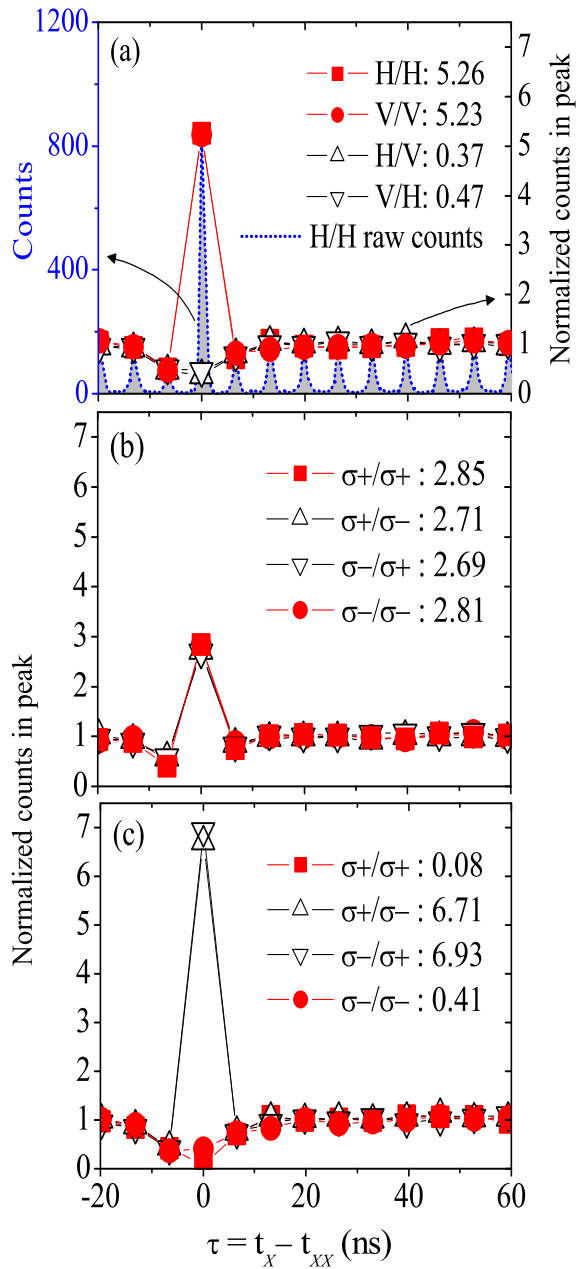}
\caption{\label{Fig:corelXX-X}(Color online) Polarized
crosscorrelation of the {\it XX} and {\it X} emission. Each panel
contains histograms for four possible polarization combinations of
the photon pair. Polarizations of correlated transitions are
indicated in {\it XX}/{\it X} order. Points represent integrated
counts in a peak, normalized to the average value for large delays.
Solid lines are guides to the eye. Magnitudes of zero delay peaks in
respective histograms are given in the each panel. Polarization
correlation of photons from {\it XX}-{\it X} cascade is evidenced in
linear H/V polarization basis at $B$~=~0~T (a) and in circular
$\sigma+/\sigma-$ polarization basis at $B$~=~5~T (c). There is no
polarization correlation in circular $\sigma+/\sigma-$ polarization
basis at $B$~=~0~T (b). Panel (a) shows additionally a raw
coincidence counts histogram (left axis).}
\end{figure}

In the experiment, the monochromators were set to detect {\it XX} and {\it X} transition by
the \textit{start} and \textit{stop} diode, respectively. The obtained correlation histograms
supplied information on relative polarizations of {\it XX} and {\it X} photons emitted
subsequently in {\it XX} radiative decay. Due to the pulsed excitation, the
histograms consist of peaks spaced equally by the repetition period of the
excitation pulses (Fig.~\ref{Fig:corelXX-X}(a)). The {\it XX}-{\it X} crosscorrelation histograms measured at $B$~=~0~T in the linear H-V polarization basis is shown in Fig.~\ref{Fig:corelXX-X}(a). The normalized areas of the central peak in Fig.~\ref{Fig:corelXX-X}(a) are $5.3~\pm~0.4$ and $5.2~\pm~0.3$ for parallel linear polarizations and $0.47~\pm~0.02$ and $0.37~\pm~0.03$ for orthogonal linear polarizations.
The measurement repeated in the rotated linear $-45^\circ/+45^\circ$
polarization basis has not revealed any polarization dependent effects on
the zero delay peak (not shown). Similarly, no polarization effects were
detected in the measurement performed in circular $\sigma+/\sigma-$ basis
(Fig.~\ref{Fig:corelXX-X}(b)). Thus, {\it XX} decay produces a pair of
classically correlated photons, which is expected for a QD with a reduced
symmetry.~\cite{Santori02, Stevenson02, Ulrich}

Degree of the correlation in linear H/V polarization basis, estimated following
Ref.~\onlinecite{Santori02}, amounts to $\chi_{HV}$~=~0.86~$\pm~0.06$. The nonzero probability of
detecting perpendicularly polarized photon pairs originates mostly from
the relaxation of excitonic spin occurring over the exciton lifetime.
Using the formula derived by Santori \textit{et al.} (Ref.~\onlinecite{Santori02}) and
basing on the {\it X} lifetime ($\tau_{radX}=~0.29 \pm 0.05$~ns) obtained from
an independent experiment performed on the same QD,~\cite{SuffPRB} we
estimate {\it X} spin relaxation time at $T_{X}$~=~$3.4 \pm 0.1$~ns. The estimated
$T_{X}$ value is an order of magnitude larger than the excitonic radiative
lifetime, in agreement with previous results obtained on InAs/GaAs~\cite{Paillard, Santori02}, CdSe/ZnSe~\cite{FlissikowskiPRL} and CdTe/ZnTe~\cite{Mackowski} QDs.

Mixing of excitonic states with angular momentum M~=~$\pm$1 decreases on application of a
magnetic field. It decreases with increasing ratio of
the Zeeman splitting to the AES.~\cite{BesombesJoCG} As the Zeeman splitting becomes dominant, the linearly
polarized excitonic doublet observed in $\mu$-PL spectra converts in two
lines with nearly perfect orthogonal circular polarizations (Fig.~\ref{Fig:PLspectrum}(b))
corresponding to almost pure M~=~$\pm$1 excitons. A simple calculation~\cite{BesombesJoCG} provides an estimate of ellipticity of the eigenstates and of the resulting circular polarization degree at 98.3\% at $B$~=~5~T. This Zeeman-controlled emission is demonstrated in polarized crosscorrelations measured at $B$~=~5~T
on spin split {\it XX} and {\it X} lines in the circular basis
(Fig.~\ref{Fig:corelXX-X}(c)). As seen in Fig.~\ref{Fig:corelXX-X}(c),
{\it XX}-{\it X} photon pairs contributing to the central peak exhibit significant,
positive (negative) correlation for opposite (equal) circular
polarizations, in contrast to the result obtained at $B$~=~0~T
(Fig.~\ref{Fig:corelXX-X}(b)). The respective normalized values of the
central peaks in histograms of Fig.~\ref{Fig:corelXX-X}(c) provide the degree of
polarization correlation~\cite{Santori02}
$\chi_{\sigma+\sigma-}$~=~0.95~$\pm~0.02$ at $B$~=~5~T. The large degree of polarization correlation
(higher than that at zero field) shows that probability of spin-flip
accompanied transition between the intermediate excitonic states of the cascade decreases when {\it X} level splitting increases in magnetic field.
Calculation of the exciton relaxation time at $B$~=~5~T (simple exciton spin-flip time in this case) made under assumption that {\it X} lifetime does not change in the magnetic field~\cite{Paillard} and including a correction for incomplete (98.3\%) circular polarization of excitonic states, yields the value $T_{X}$~=~16~$\pm~3$~ns. This is over four times larger than the value determined for the zero field case. The obtained value will be used as a
parameter in the rate equation model introduced in Sec.~\ref{sec:Model}.

In summary, the set of {\it XX}-{\it X} crosscorrelations measured
(Fig.~\ref{Fig:corelXX-X}) shows that pairs of photons emitted from an
anisotropic QD in {\it XX}-{\it X} cascade exhibit at $B$~=~0~T only a strong classical correlation
in the linear polarization basis corresponding to symmetry axes of the
dot, in agreement with previous experiments.~\cite{Santori02, Stevenson02,
Ulrich} Anisotropy induced collinear polarization correlation of photons
emitted in {\it XX}-{\it X} cascade is converted to a counter-circular polarization
correlation after applying magnetic field parallel to the sample growth
axis. Exciton spin relaxation is found to be slowed down in the presence of
magnetic field, as demonstrated by increase of the {\it X} spin relaxation time
from $3.4~\pm~0.1$~ns at $B$~=~0~T to $16~\pm~3$~ns at $B$~=~5~T.

\subsection{\label{sec:Results.3} Single carrier spin memory effects}
Previous investigations of QD emission by polarization resolved
correlation spectroscopy have been limited to the {\it XX}-{\it X} cascade, which was
found to produce polarization-correlated~\cite{Santori02, Stevenson02,
Ulrich} or polarization-entangled~\cite{StevensonNature, Young, Akopian}
triggered photon pairs. In this section, we present results of time and
polarization resolved correlations between charged exciton and neutral
exciton photons emitted from a single CdTe/ZnTe QD. We examine the
influence of magnetic field on the carrier spin dynamics. The measurements
revealed long lasting carrier spin memory in magnetic field and confirmed an effective carrier spin \textit{read-out}.

\begin{figure}
\includegraphics[width=80mm]{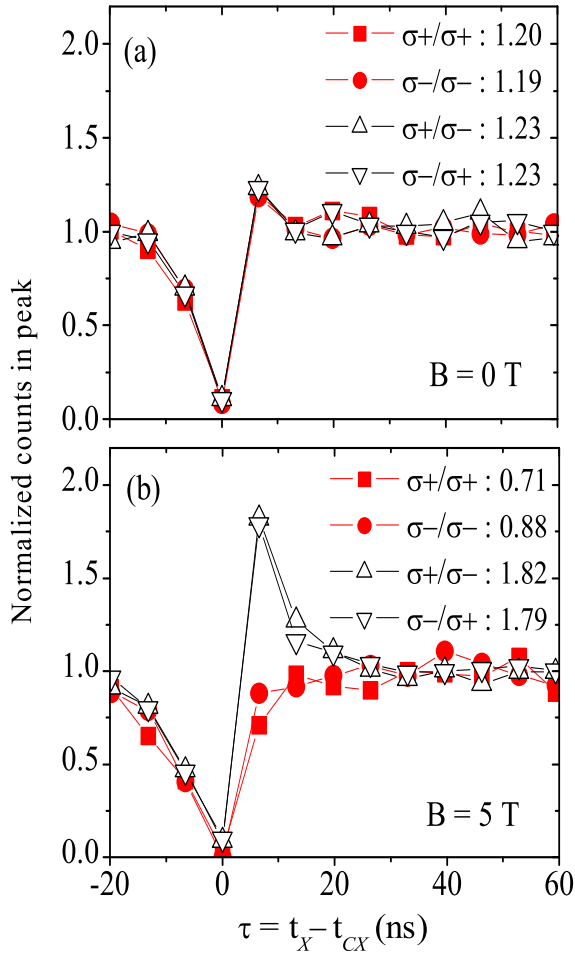}
\caption{\label{Fig:corelCX-X}(Color online) Histograms of {\it
CX}-{\it X} crosscorrelation measured in the circular
$\sigma+/\sigma-$ polarization basis for four possible combinations
of the polarization of the {\it CX}-{\it X} photon pair (a) at
$B$~=~0~T and at (b) $B$~=~5~T. Polarizations of correlated
transitions are indicated in {\it CX}/{\it X} order. Points
represent integrated and normalized (to the average at large delays)
number of counts in a peak. Solid lines are guides to the eye.
Values given in panels represent normalized number of counts in
$m$~=~1 peak of a respective histogram. Excitation power at
$I_{XX}/I_{X}=0.30$ (see Sec.~\ref{sec:Model}). Enhanced $m$~=~1
peaks in the case of correlation between orthogonal {\it CX} and
{\it X} polarizations indicate transfer of spin orientation from
{\it CX} to {\it X} over time of the repetition period (b).}
\end{figure}

The measurements were performed in the linear or in the circular polarization
basis. The results of {\it CX}-{\it X} crosscorrelation involving {\it CX} and {\it X} emission
measured at $B$~=~0~T in the circular polarization basis are presented in Fig.~\ref{Fig:corelCX-X}(a). All the
histograms in Fig.~\ref{Fig:corelCX-X}(a) have their central peak strongly
suppressed and exhibit an asymmetric shape, characteristic for
{\it CX}-{\it X} crosscorrelation. This is known to
originate from the QD charge state variation under nonresonant excitation,
which favors capture of single carriers instead of entire excitons.~\cite{SuffPRB} The central peak of the {\it CX}-{\it X} histogram (Fig.~\ref{Fig:corelCX-X}) represents the detection of pairs
consisting of {\it CX} and {\it X} photons emitted following the same excitation
pulse, therefore its suppression reflects expected antibunching of {\it CX} and
{\it X} photons. Peaks at a negative (positive) delay represent pairs of
photons detected following different pulses, such that {\it X} photon precedes
(succeeds) {\it CX} photon.
The dependence of circular polarization of {\it X} emission on circular
polarization of previously emitted {\it CX} photon ($m > 0$ peaks) would mean
that the spin orientation of the {\it CX} that recombined affects the spin
orientation of subsequently formed {\it X}. In such a case the carrier present in
the dot after {\it CX} recombination would provide transfer of the spin polarization from {\it CX} to {\it X}. Its spin state would be \textit{read} from the polarization of {\it X} emission. However, no dependence of peak intensity on combination of photon pair polarizations in neither circular (Fig.~\ref{Fig:corelCX-X}(a)) nor linear H/V
(not shown) polarization bases is observed at $B$~=~0~T. We deduce therefore
that polarization of the carrier left in the dot after {\it CX} recombination is not transferred to the {\it X} photon subsequently emitted by the QD. This may be caused by the anisotropic exchange splitting of the {\it X} state, resulting in the averaging out of the circular polarization by precession between two linearly polarized eigenstates. For the same reason, no optical orientation of excitons is observed in anisotropic quantum dots.~\cite{KowalikAPL07}

However, the polarization transfer becomes significant on application of magnetic field, when both {\it CX} and {\it X} emit in common circular $\sigma+/\sigma-$ polarization basis. At $B$~=~5~T peaks at small positive delays show a significant enhancement or suppression for opposite or equal circular
polarizations, respectively (Fig.~\ref{Fig:corelCX-X}(b)). The
first peak at positive delay ($m$ = 1 peak), represents {\it X}
photon detection in the pulse immediately following the detection of {\it CX}
photon. Its normalized areas are $1.8 \pm 0.1$ and $1.8 \pm 0.1$ for
photons of opposite polarization and $0.88 \pm 0.05$ and $0.71 \pm 0.05$
for photons of the same polarization. This is an evidence of spin memory
in magnetic field.

The intensities of the $m$ = 1 peaks in Fig.~\ref{Fig:corelCX-X}(b) correspond to polarization degree of 39\%.
Further ($m > 1$) peaks of the {\it CX}-{\it X} histogram also
show a polarization, which decreases with increasing peak number.
This occurs because each additional excitation pulse reduces the probability that the
dot remains in the original, \textit{post} {\it CX} recombination, single carrier spin state.

In summary, the results of {\it CX}-{\it X} crosscorrelation in magnetic field
provide a clear evidence of the polarization memory extending over a few excitation pulses and of effective optical \textit{read-out} of the single carrier spin in the dot.

\section{\label{sec:Model}Model description of the polarized {\it CX}-{\it X} crosscorrelation}
As already mentioned, the singlet ground state of the charged exciton
contains two identical carriers of opposite spins. One of them
decays in {\it CX} recombination, emitting a photon with circular
polarization determined by its spin. The spin polarization of the
second carrier will determine the polarization of an {\it X}
photon emitted during {\it X} recombination after next laser pulse. This will happen after trapping a carrier of opposite charge. If the spin polarization is conserved
over the repetition period, the {\it CX} and {\it X}
photons produced by two consecutive pulses will thus have opposite
circular polarizations. In reality, this spin conservation is never
perfect and can be measured by polarization correlation coefficient
$P$ defined as
\begin{equation}
\label{Eq:Pexp} P=\frac
{I_{\sigma-/\sigma+}+I_{\sigma+/\sigma-}-I_{\sigma+/\sigma+}-I_{\sigma-/\sigma-}}
{I_{\sigma-/\sigma+}+I_{\sigma+/\sigma-}+I_{\sigma+/\sigma+}+I_{\sigma-/\sigma-}}
\end{equation}
where $I_{\gamma/\delta}$ denotes intensity of the correlated counts
in the {\it CX}-{\it X} histogram measured with {\it CX} and {\it X} photons detected in
polarizations $\gamma$ and $\delta$, respectively.

\begin{figure}
\includegraphics[width = 84mm]{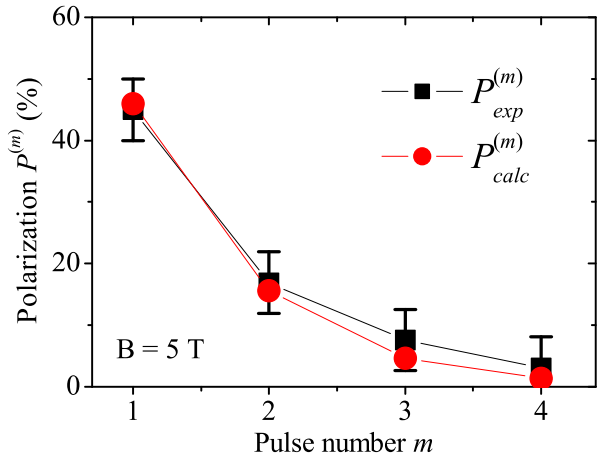}
\caption{\label{Fig:PolDegree}(Color online) Polarization
correlation versus pulse number at $B$~=~5~T. Experimental values
(squares) are compared with calculation (dots). Model parameters
$\alpha~=~0.71$, $\beta~=~0.66$, $\xi~=~0.22$, $\kappa~=~0.69$ (see
text). Excitation power at $I_{XX}/I_{X}=0.26$. Lines added to guide
the eye.}
\end{figure}

In the experiment we measure $P_{exp}^{(m)}$ values related to consecutive
peaks of the {\it CX}-{\it X} histogram, expressed by total correlated counts $I_{\gamma/\delta}^{(m)}$ of respective peaks, numbered by index $m$. Values of $P_{exp}^{(m)}$ determined for peaks of $1 \leq m \leq 4$ are shown in Figure~\ref{Fig:PolDegree}.
The coefficient $P^{(m)}$ is equivalent to the ratio of the probability that {\it X} photon with polarization defined by the carrier spin conservation is emitted following the $m$-th pulse to the overall probability of {\it X} photon emission following the $m$-th pulse. Thus, $P^{(m)}$ can be written taking into account different polarization loss mechanisms. E.g., for $m$~=~1 peak:
\begin{equation}
\label{Eq:Pcalc}
P_{calc}^{(1)}=\;\epsilon \; \frac{n_{X}^{(1)}}{n_{X}^{(1)}+n_{XX}^{(1)}} \; \frac{T_{X}}{T_{X}+\tau_{radX}}\; \kappa
\end{equation}
where symbols $n_{i}^{(m)}$ denote occupation probability of QD
states directly after an excitation pulse, which is assumed to be
short enough to neglect recombination during excitation (see
discussion further below). The upper index given in parentheses
encodes the pulse number $m$, while the lower index $i$ encodes the
dot state. Factor $\epsilon$ (in the case of selected QD estimated
at 96.2\% at $B$~=~5~T) represents the impact of elliptical
polarization of the excitonic eigenstate. The fraction with {\it XX}
and {\it X} occupation probabilities represents polarization loss
due to biexciton recombination channel, where no polarization is
transferred by the singlet {\it XX} state. Since $T_{X}$ is {\it X}
spin-flip time (see Sec.~\ref{sec:Results.2}), expression
$T_{X}$/($T_{X} + \tau_{radX}$) represents the loss of the exciton
spin polarization during its lifetime. Finally, $\kappa$ represents
other possible loss mechanisms. In particular, it could be a
spin-flip of the remaining carrier. However, we checked that magnitude of the peaks in the polarized {\it
CX}-{\it X} correlation histogram remains unchanged after doubling
of the excitation repetition period (not shown). This indicates a
negligible carrier spin relaxation over the timescale comparable
with the repetition period. A possible polarization loss during {\it
X} state formation by capture of a second carrier of an opposite
charge will be also found negligible (see discussion in the
following). Hence, the respective polarization loss results from an
interaction of the confined electron with non-equilibrium population of carriers (exchange interaction) and/or phonons (possible local heating of the sample) excited by laser pulses.

The timescale of the observed memory effect might
also suggest that {\it CX} recombination leaves in the dot an
electron and not a hole. In contrast to the case of the electron,
the spin relaxation of the hole is known to be relatively
fast,~\cite{Flissikowski} occurring in the time of the order
comparable with the repetition period of probing the carrier's spin
state in our experiment. Thus, we make tentative assumption that
{\it CX} is negatively charged.

We shall also comment on the influence of dark exciton formation on excitonic polarization degree. Dark, nonradiative exciton state is formed in the case when carrier is captured by the dot already containing a carrier of opposite charge and the same spin. After a spin-flip process the dark exciton converts into a bright state and decays radiatively. Such an excitonic luminescence could lower the effective excitonic PL polarization degree. However, the time constant of dark exciton spin-flip is large compared to the excitation repetition period. This is known from the comparison between the results of unpolarized {\it CX}-{\it X} crosscorrelation measurements performed with two different repetition periods.~\cite{SuffPRB} They reveal no significant difference in intensity of peaks of the same number in two histograms. Thus, the influence of the dark exciton formation on exciton polarization degree can be neglected and it was not taken into account in the construction of the model describing experimental data.

The measured polarization may be understood as a product of carrier spin polarization reduced by \textit{read-out} efficiency (first two factors in Eq.~\ref{Eq:Pcalc}) and loss mechanisms (last two factors in Eq.~\ref{Eq:Pcalc}). In order to determine the most prominent factors lowering the measured polarization we introduce a simple rate equation model. It allows us to compute the occupation probabilities necessary for calculation of $P_{calc}^{(m)}$. We consider a ladder of states involving five states: from the empty dot to the biexciton state. Since only one charged exciton line of significant intensity is observed, we neglect  states of opposite charge (corresponding weak trion line visible in Fig.~\ref{Fig:PLspectrum} at 2217~meV). We define a set of variables $n_{0}^{(m)}$, $n_{e}^{(m)}$, $n_{X}^{(m)}$, $n_{CX}^{(m)}$, $n_{XX}^{(m)}$ describing the occupation of levels (encoded by the lower index) just after the excitation by the $m$-th pulse is finished. It is known from independent measurements on the same sample~\cite{Korona} that the effective excitation pulse duration ($\sim20$~ps) is much shorter than radiative decay times (hundreds of ps), therefore we neglect recombination during the excitation process. Excitation through capture of single carriers of both signs and entire excitons with respective time dependent rates $\alpha(t) = \alpha{}\cdot{}f(t)$, $\beta(t) = \beta{}\cdot{}f(t)$, and $\xi(t) = \xi{}\cdot{}f(t)$, is assumed. A common normalized excitation pulse shape $f(t)$ is assumed ($\int_{0}^{T_{rep}} f(t){\rm d}t = 1$), while $\alpha$, $\beta$, $\xi$ represent time-integrated capture rates per pulse. Escape of the carriers out of the dot is not taken into account as it has been shown to be negligible.~\cite{SuffPRB} Our simulations show that within the assumptions of the model, the shape $f(t)$ of the excitation pulse is not important. We describe the excitation process with the following set of rate equations:
\begin{subequations}
\label{Eq:RateEq}
\begin{equation}
\frac {{\rm d}n_0^{(m)}}{{\rm d}t} = - (\alpha(t) + \xi(t)) n_0^{(m)}
\end{equation}
\begin{equation}
\frac {{\rm d}n_e^{(m)}}{{\rm d}t} = \alpha(t) n_0^{(m)} -
(\beta(t)+ \xi(t))n_e^{(m)}
\end{equation}
\begin{equation}
\frac {{\rm d}n_X^{(m)}}{{\rm d}t} = \xi(t) n_0^{(m)} + \beta(t)
n_e^{(m)} - (\alpha(t) + \xi(t)) n_X^{(m)}
\end{equation}
\begin{equation}
\frac {{\rm d}n_{CX}^{(m)}}{{\rm d}t} = \xi(t) n_e^{(m)} + \alpha(t)
n_X^{(m)}-\beta(t) n_{CX}^{(m)}
\end{equation}
\begin{equation}
\frac {{\rm d}n_{XX}^{(m)}}{{\rm d}t} = \xi(t) n_X^{(m)} + \beta(t)
n_{CX}^{(m)}
\end{equation}
\end{subequations}
We assume purely radiative decay of excitons. Thus, the initial
conditions for a consecutive excitation pulse are determined by the
final state of excitonic recombination after preceding excitation
pulse. For the particular case of $m~=~1$, the initial conditions
are in a good approximation $n_{e}^{(1)}(0)~=~1$ and $n_{i\neq
e}^{(1)}(0)~=~0$ (single carrier left after {\it CX} recombination
present in the dot). Integrated capture rates $\alpha$, $\beta$,
$\xi$ obtained from the experiment on the same QD with no
polarization resolution were used.~\cite{SuffPRB} They were scaled
by a common factor in order to take into account a variable
excitation power. The factor was adjusted to fit the ratio of {\it
XX} to {\it X} emission intensity ($I_{XX}/I_{X}$). (Consistently
with Ref.~\onlinecite{SuffPRB}, $\alpha$ /$\beta$/ represents
capture rate of the first /the second/ carrier to the dot, that is
electron /hole/.)

\begin{figure}
\includegraphics[width = 74mm]{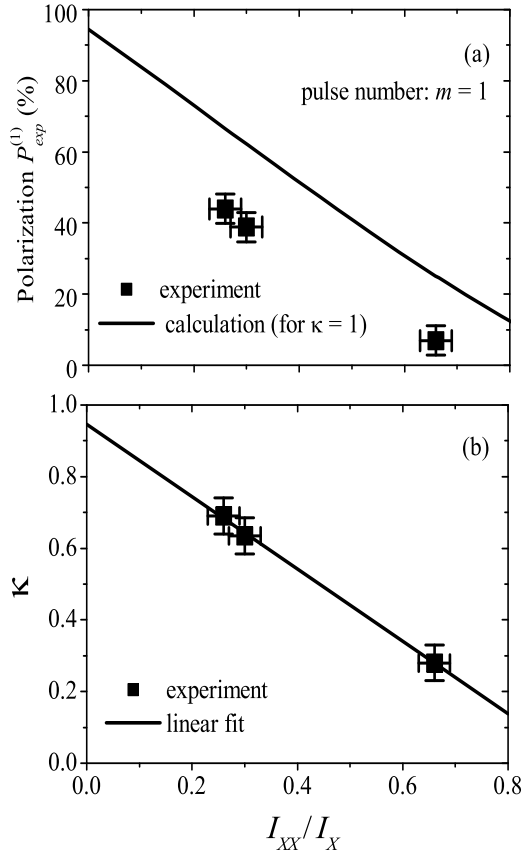}
\caption{\label{Fig:Pcalc}(a) Polarization correlation at $B$~=~5~T for the first pulse after {\it CX} emission $P_{exp}^{(1)}$ versus excitation power represented by the ratio $I_{XX}/I_{X}$. Solid line -- calculation ($\kappa = 1$) according to Eq.~\ref{Eq:Pcalc}. (b) Coefficient $\kappa$ of carrier spin conservation at $B$~=~5~T determined from fitting of the $P_{calc}^{(1)}$ to $P_{exp}^{(1)}$ for different excitation powers.}
\end{figure}

The first consequence of the model is the variation of the calculated polarization coefficient $P$ with the excitation power. Also the $P_{exp}^{(1)}$ decreases with the increasing excitation power. This is expected, since the contribution of {\it X} photons coming from {\it XX} radiative decay increases with the excitation intensity. They are effectively unpolarized and they lower the value of $P_{exp}^{(m)}$. We determined $P_{calc}^{(1)}$ for different excitation powers by solving Eqs.~\ref{Eq:RateEq} with suitably scaled rates $\alpha$, $\beta$, and $\xi$ assuming full conservation of the electron spin ($\kappa~=~1$). The Figure~\ref{Fig:Pcalc}(a) shows comparison of $P_{calc}^{(1)}$ for $\kappa~=~1$ and $P_{exp}^{(1)}$ plotted as a function of the $I_{XX}/I_{X}$ ratio. The ratio $I_{XX}/I_{X}$ was chosen to represent excitation power, since it provides a convenient measure of excitation intensity.
The discrepancy between the experimental and the calculated values is a clear indication that conservation of the electron spin polarization between the {\it CX} and {\it X} emissions is not perfect.

Thus, we fitted $P_{calc}^{(1)}$ to $P_{exp}^{(1)}$ with $\kappa$ being the (only) fitting parameter. Values of $\kappa$ determined this way are shown in Fig.~\ref{Fig:Pcalc}(b) as a function of $I_{XX}/I_{X}$ ratio. As visible in the Fig.~\ref{Fig:Pcalc}(b), $\kappa$ attains the value of $\kappa = 0.69$ at $I_{XX}/I_{X}~=~0.26$ and decreases with the increasing excitation power. A linear fit to the experimental points is also shown in the Fig.~\ref{Fig:Pcalc}(b). The electron polarization conservation, and the electron spin optical \textit{read-out} are almost perfect in the limit of low excitation power.

This means that the capture of the hole to form an exciton with the electron in the QD does not induce any significant polarization loss, since process of {\it X} formation does not depend on excitation power. The dependence of $\kappa$ on the excitation intensity confirms that $\kappa$ originates from the factors such as interaction of the confined electron with carriers and/or phonons generated by the excitation pulse, as contribution of these factors depends on the excitation power.

Therefore, the loss of the electron polarization represented by $\kappa$ takes place following each excitation pulse. This allows us to write down $P_{calc}^{(m)}$ for consecutive pulses in the form:
\begin{equation}
\label{Eq:Pcalcm}
P_{calc}^{(m)} = (n_{e}^{(1)})^{m-1}\;\epsilon\;\frac{n_{X}^{(1)}}{n_{X}^{(m)}+n_{XX}^{(m)}}\;\frac{T_{X}}{T_{X}+\tau_{radX}}\;\kappa^m
\end{equation}
The factor $(n_{e}^{(1)})^{m-1}$ reflects the probability that QD keeps its one-carrier state unchanged through $m-1$ excitation pulses. The factor $\kappa^m$ represents degree of conservation of the electron spin following $m$ consecutive excitation pulses. Its exponential dependence on the peak number $m$ (evidenced by Fig.~\ref{Fig:PolDegree}) confirms the role of laser pulses as the source of polarization loss described by $\kappa$.

Integration of Eqs.~\ref{Eq:RateEq} yielded probabilities $n_{i}^{(m)}$ of finding the dot in a state $i$ following $m$-th excitation pulse. Thus $P_{calc}^{(m)}$ for consecutive pulses were calculated from Eq.~\ref{Eq:Pcalcm}. They are compared with the experimental values of $P_{exp}^{(m)}$ for peaks of $1\leq~m\leq~4$ in Fig.~\ref{Fig:PolDegree} for the example case of $I_{XX}/I_{X}~=~0.26$. The satisfactory agreement between calculated and experimental values justifies the introduced model.

To summarize this Section, {\it CX}-{\it X} crosscorrelation measurements provide evidence for single carrier spin polarization memory in magnetic field. We described quantitatively the polarization memory after consecutive excitation pulses. Comparison between the model and the experiment shows that carrier spin conservation and the optical \textit{read-out} efficiency are close to 100\% in the limit of low excitation power. The maximum efficiency of the spin \textit{read-out} obtained in the experiment is 68\%. The {\it CX}-{\it X} crosscorrelation performed for different excitation powers allowed us to verify the expected influence of biexciton formation on the loss of polarization memory.
\section{Conclusions}
We performed polarized crosscorrelation measurements of photons from exciton, biexciton, and trion recombination in a single, anisotropic CdTe/ZnTe quantum dot. In absence of magnetic field we observed a
strong collinear polarization correlation ($\chi_{HV} = 0.86 \pm 0.06$) of photons
emitted in the biexciton-exciton cascade. When Zeeman splitting dominates over anisotropic exchange splitting, the photons in the cascade become correlated in opposite circular polarizations ($\chi_{\sigma+\sigma-} = 0.95 \pm 0.02$). Exciton spin relaxation time values $T_{X}$~=~$3.4 \pm 0.1$~ns at $B$~=~0~T and $16 \pm 3$~ns at $B$~=~5~T were determined from the {\it XX}-{\it X} correlation measurements.

Trion-exciton crosscorrelation measurements conducted in magnetic field of 5~T
have revealed long timescale (at least tens of ns) polarization
memory in the QD excitonic emission. Effective optical \textit{read-out} of the spin polarization of a single carrier confined in the anisotropic QD has been demonstrated in magnetic field. The decay of the polarization memory with the increasing number of excitation pulses separating two correlated photons has been described with a simple rate equation model. It was attributed to the combined influence of competitive biexciton spin singlet recombination and loss of carrier spin polarization, perturbed primarily by the very laser light used for the spin \textit{read-out}. Efficiency of optical \textit{read-out} of the spin in magnetic field turned out to be dependent on the excitation power (the maximum efficiency obtained in the experiment was 68\%).

The results obtained indicate CdTe/ZnTe QDs as a valuable proving ground for future applications of polarization controlled single photon emitters or spin qubits in the area of quantum information processing.

\begin{acknowledgments}
This work was partially supported by the Polish Ministry of Science and Higher
Education research grants in years 2005-2010 and by European project no. MTKD-CT-2005-029671.
\end{acknowledgments}


\end{document}